\documentclass{emulateapj}

\def\ptsec{$^{\prime\prime}\mskip-7.6mu.\,$}

\shorttitle{Physical Conditions of Accreting Gas}
\shortauthors{Bary et al.}

\begin{document}

\title{Physical Conditions of Accreting Gas in T Tauri Star Systems}

\author{Jeffrey S. Bary\altaffilmark{1}, Sean P. Matt\altaffilmark{1}, Michael F. Skrutskie\altaffilmark{1}, John C. Wilson\altaffilmark{1}, Dawn E. Peterson\altaffilmark{2}, and Matthew J. Nelson\altaffilmark{1}} 
\email{jsb3r@virginia.edu, spm5x@virginia.edu, mfs4n@virginia.edu, jcw6z@virginia.edu, dpeterson@cfa.harvard.edu, mjn4n@virginia.edu}
\altaffiltext{1}{Department of Astronomy, University of Virginia, P.O. Box 400325 
Charlottesville, VA 22904-4325}
\altaffiltext{2}{Harvard-Smithsonian, Center for Astrophysics, 60 Garden St., Cambridge, MA  02138}

\begin{abstract}

We present results from a low resolution (R~$\simeq$~300)
near-infrared spectroscopic variability survey of actively accreting T Tauri stars
(TTS) in the Taurus-Auriga star forming region.  Paschen and
Brackett series H~{\scshape i} recombination lines were detected in
73 spectra of 15 classical T Tauri systems.  The values of the
Pa$n_{up}$/Pa$\beta$, Br$n_{up}$/Br$\gamma$, and Br$\gamma$/Pa$n_{up}$
H~{\scshape i} line ratios for all observations exhibit a scatter of~$\la$~20\% about the weighted mean, not only from source to source, but also for epoch-to-epoch variations in the same source.  A representative or `global' value was determined for each ratio in both the Paschen and Brackett series as well as the Br$\gamma$/Pa$n_{up}$ line ratios.  A comparison of observed line ratio values was made to those predicted by the temperature and electron density dependent models of Case~B hydrogen recombination line theory.  The measured line ratios are statistically well-fit by a tightly constrained range of temperatures ($T$~$\la$~2000~K) and electron densities  (10$^{9}$~$<$~n$_{\rm e}$~$\la$~10$^{10}$~cm$^{-3}$).  A comparison of the observed line ratio values to the values predicted by the optically thick and thin local thermodynamic equilibrium cases rules out these conditions for the emitting H~{\scshape i} gas.  Therefore, the emission is consistent with having an origin in a non-LTE recombining gas.  While the range of electron densities is consistent with the gas densities predicted by existing magnetospheric accretion models, the temperature range constrained by the Case~B comparison is considerably lower than that expected for accreting gas.  The cooler gas temperatures will require a non-thermal excitation process (e.g., coronal/accretion-related X-rays and UV photons) to power the observed line emission.

\end{abstract}


\section{Introduction}
\label{intro}

In the decades since their discovery, photometric variability continues to be a defining characteristic of T Tauri stars (TTS) and has
been observed at most wavelengths with timescales as short as
hours \citep[e.g.,][]{joy1945,bouv1993}.  In addition to the widely studied photometric variations of TTS, astronomers have begun, more recently, to probe the variability of the unique spectral emission features associated with these sources using multi-epoch spectroscopic observations to note variations in the
shapes and line strengths of the emission features
\citep[e.g.,][]{john1995,basr1997,alen2001,bouv2007}.  An investigation of the physical conditions that give rise to these emission features will illuminate many of the processes involved in the formation of these stars.

Single-epoch observations of H~{\scshape i} Balmer series line
emission from TTS reveal a complicated kinematic line structure, with
features often including blue-shifted and/or red-shifted absorption components.  As
evident from the catalog of H$\alpha$ features presented in
\cite{reip1996}, a variety of complex profiles exist for this feature
and several other Balmer emission lines.  The same spectral line
studies reveal quite large full-width at half of maxima (FWHM) for the
emission components of these features and suggest that the gas is
moving at hundreds of km~s$^{-1}$ relative to the sources.  The complexity of these emission features alone has led to different conclusions about the
origins of this emitting gas.  However, the blue-shifted absorption
components or P Cygni profiles for these emission features are most likely produced by gas entrained in an outflowing wind \citep{hart1990,grin1991,calv1992}.  As a result, the
observation of a large H$\alpha$ equivalent width (EW) for an
individual source was initially interpreted as evidence of and arising entirely from an active outflowing wind.

In another model of the star-disk interaction, the H~{\scshape i} lines and other TTS phenomena were formed in a viscous boundary layer, where the accretion disk touched the surface of the star \citep{lynd1974,bert1989}.  However, the inability of both the boundary layer and stellar wind models to explain the shapes and widths of H~{\scshape i} emission features led to the application of the magnetospheric accretion model to TTS \citep{ghos1979,koen1991}.  Previously developed to explain disk accretion onto compact objects, the magnetospheric accretion model established a new paradigm for the star-disk interaction in these pre-main sequence stars.  This model involves a stellar magnetosphere that truncates the disk, opening a gap with a size of a few stellar radii between the disk and the surface of the star.  Looping stellar magnetic field lines thread the inner most regions of these disks, acting as accretion funnels through which disk matter may free fall onto the stellar surface.  The free falling material, confined to a magnetic channel, naturally explains the large velocities observed for the H~{\scshape i} emission features, as well as the appearance of red-shifted absorption components.  In addition, the accretion shocks formed at the point of impact at the stellar surface provide the heated gas necessary to produce the `blue' continuum excesses characteristic of actively accreting TTS.  Several authors have employed the magnetospheric accretion model to predict the fluxes and shapes of a few of the optical and infrared H~{\scshape i} emission features \citep{hart1994,muze1998a,muze1998b}.  These modeling efforts support the idea that the bulk of the emitting H~{\scshape i} gas is confined to the columns of accreting gas, rather than an outflowing wind.

Advances in infrared spectroscopy have allowed astronomers to observe
at high spectral resolution a number of H~{\scshape i} Paschen and
Balmer series emission lines, predominantly Pa$\beta$ and Br$\gamma$,
from TTS \citep{naji1996,muze1998a,muze1998b,folh1998,folh2001,natt2006}.
These infrared H~{\scshape i} lines, which possess lower opacities relative to the Balmer series, `conspicuously' lack the blue-shifted absorption features present in
H$\alpha$ and H$\beta$ profiles \citep{folh1998}.  The infrared lines
also appear to correlate with the mass accretion rates
determined from measurements of the `blue' continuum excess
\citep{gull1998,gull2000,muze2001} suggesting that the H~{\scshape i}
emission is related to accretion activity.  Therefore, as a result of the
successes of the magnetospheric accretion model in explaining the
shapes of the H~{\scshape i} emission features and the correlation of these features with other accretion diagnostics, the interpretation of the H~{\scshape i} emission lines and the location of the emitting gas has shifted in the last 15 years.  Where once the presence of a strong H$\alpha$ emission line in the spectrum of a TTS was widely interpreted as an indicator of an outflowing wind and the `blue' continuum excesses were evidence of a hot boundary layer, today the H~{\scshape i} features and the `blue' excesses are both interpreted as evidence of magnetospheric accretion.

More recently, spectro-astrometric observations of several classical TTS (cTTS; TTS possessing evidence for actively accreting circumstellar disks) and Herbig AeBe stars possessing symmetric (Type~I) Pa$\beta$ emission features have shown that some of the emission in the line-wings have measurable offsets with respect to the central sources.  These results suggest that outflowing gas is a likely source for this component of the emission \citep{whel2004} and potentially explain the difficulty \cite{folh2001} encountered modeling the line-wings of these emission features with a magnetospheric accretion scenario that did not include a wind.  In addition, AMBER/VLTI observations of Herbig Ae star HD~104237 revealed that the Br$\gamma$ and {\it K}-band continuum emission have similar spatial distributions \citep{tatu2007}.  \cite{tatu2007} successfully explain the visibility of the Br$\gamma$ emission by modeling it with an outflowing wind and places the emission at the base of the jet with a radial extent of 0.2 to 0.5~AU.  Observations such as these indicate that a portion of the near-infrared H~{\scshape i} emission features can be produced in outflowing gas. 

Despite these apparent advancements in understanding the origin of the H~{\scshape i} emission features, the physical conditions of the emitting gas (e.g., ionization state, level populations, temperature, density) remain poorly constrained \citep{alen2000,whel2004,tatu2007,eisn2007a}.  \cite{mart1996} calculated a temperature and density profile for gas traveling along an accretion funnel from the disk to the star by modeling several sources of gas heating (e.g., ambipolar diffusion, adiabatic compression, photoionization, and external radiative excitation) and cooling (e.g., Ca {\scshape ii} and Mg {\scshape ii} ions).  Assuming an initial gas temperature of 3000~K for gas located at the inner edge of the disk where it is loaded into the accretion funnel, for a fiducial mass accretion rate of $\dot{M}$~=~10$^{-7}$~M$_\odot$~yr$^{-1}$, \cite{mart1996} found that the gas temperature profile peaks at $T$~$\sim$~6500~K.

In another investigation seeking to constrain the temperature of magnetospherically
accreting gas, \cite{muze2001} calculated H$\alpha$ and H$\beta$ line fluxes and
profiles dependent upon $\dot{M}$ and a range of gas temperatures.  The authors found `optimal' temperatures of 6000~$<$~$T$~$<$~12000~K for a range of accretion rates 10$^{-6}$~$\ge$~$\dot{M}$~$\ge$~10$^{-10}$~M$_\odot$~yr$^{-1}$, with the lower gas temperatures corresponding to sources with higher mass
accretion rates.  This study suggested that the gas temperatures predicted by
\cite{mart1996} were too low to produce the observed line luminosities
for the $\dot{M}$ values considered.  These temperatures agree with
the standard thermal gas temperature of $T$~$\sim$~10$^4$~K observed for
most H~{\scshape ii} emission regions.

A direct measurement of the physical conditions of the gas will provide a valuable test of the models and may help to further constrain the origin of the H~{\scshape i} line emission.  In optically thin gases, line ratios provide valuable information about the physical conditions.  However, at high densities, optical depth effects require radiative transfer modeling and thus complicate the interpretation of these line measurements.  In cTTS, H$\alpha$ emission lines often appear to be optically thick and may be formed by the contributions of emitting gas in multiple regions of the star-disk system, not limited to the accretion flow alone \citep{kuro2006}.  The near-infrared H~{\scshape i} lines, with higher $n$-values, have lower optical depths and appear to be less complicated than the Balmer transitions \citep{folh2001}. Observational attempts to compare the measured values of the Pa$\beta$/Br$\gamma$ H~{\scshape i} infrared line ratio to the values predicted for optically thin and thick hydrogen gas in LTE, in the past, have failed to determine a temperature for the emitting gas \citep{muze2001}.  Similarly, comparisons of the Pa$\beta$/Br$\gamma$ ratio to the values predicted using the Case~B approximation of hydrogen recombination line theory (\cite{stor1995} and references therein) have also failed to find a meaningful result.  In fact, the lack of agreement between the observed and predicted line ratios lead to the conclusion that the emitting H~{\scshape i} gas is neither optically thin or thick nor can it be described as a recombining gas \cite{muze2001}.  However, in most cases where comparisons were made, the values for this line ratio were determined from spectra collected at different times.  Given that the variability of infrared H~{\scshape i} line fluxes has timescales as short as hours \citep{john1995b,alen2001,bouv2007}, a line ratio calculated from two distinct, non-simultaneous spectral observations will likely be misleading.  Therefore, it is not surprising that previous comparisons of a single line ratio have failed to meaningfully constrain the physical conditions of the emitting gas.  Simultaneous observations of not only Pa$\beta$ and Br$\gamma$ \citep[e.g.,][]{gatt2006}, but including several other Paschen and Brackett series emission lines will improve upon these previous comparisons.

We present an analysis of H~{\scshape i} spectral line data {\it simultaneously} acquired as part of a multi-epoch near-infrared spectroscopic variability study of cTTS in the Taurus-Auriga star forming region.  Multiple Paschen and Brackett emission features were detected in the spectra of our sample of actively accreting cTTS collected during the three year survey.  We determine the values for 16 distinct line ratios that are representative of the whole sample of TTS.  In \S\ref{analysis}, we compare the set of observed ratios, broken into three groups, Paschen series, Brackett series, and Br$\gamma$/Pa$n_{up}$, where $n_{up}$ is the upper level of the transition, with the temperature and electron density sensitive Case~B models as well as the temperature-only sensitive optically thick and thin LTE scenarios.  The results of these comparisons and their predictions of the physical conditions of the emitting H~{\scshape i} gas are discussed in \S\ref{disc}.

\section{Observations}
\label{obs}

Multi-epoch spectroscopic observations were made of 16 cTTS in the Taurus-Auriga star forming region.  The spectra were acquired using CorMass (Cornell-Massachusetts), a cross-dispersed low resolution (R~$\sim$~300) spectrometer with continuous coverage from 0.8 to 2.5~$\mu$m \citep{wils2001}, with the 1.8-m Vatican Advanced Technology Telescope (VATT) atop Mt.\ Graham in Arizona.  The wavelength coverage of CorMass permits simultaneous observation of multiple Paschen and Brackett series H~{\scshape i} emission lines as well as He {\scshape i} (1.083~$\mu$m) and the Ca~{\scshape ii} infrared triplet (0.8498, 0.8542, and 0.8662~$\mu$m).  A total of 104 spectra of 16 targets were obtained during the period of December 2003 to January 2005.  The cTTSs observed, along with previously measured values for H$\alpha$ EWs, $\dot{M}$, and X-ray luminosities are listed in Table~1.

\begin{deluxetable}{lccc}
\tablewidth{0pt}
\tablecaption{Target Star Data \label{tab_targets}}
\tablehead{
\colhead{Star Name} &
\colhead{H$\alpha$ EW$^a$} &
\colhead{$\dot M$ $^b$} &
\colhead{$L_X$$^c$} \\
\colhead{} &
\colhead{(\AA)} &
\colhead{($10^{-7} M_\odot$ yr$^{-1}$)} &
\colhead{($10^{30}$ erg s$^{-1}$)}
}

\startdata

AA Tau & 37--46      &  0.04    & 1.039    \\
BP Tau & 40--92      & 0.23     & 1.482    \\
CW Tau & 135--140 & 1.27    & \nodata   \\
CY Tau & 55--70      & 0.08    & 0.108, 0.202    \\
DF Tau & \nodata    & 2.08    &  \nodata       \\
DG Tau & 63--125   & 9.19     & 0.252        \\
DM Tau & 139         & \nodata & 0.181        \\
DO Tau & 100         & \nodata & \nodata      \\
DQ Tau & \nodata   & \nodata &  \nodata     \\
DR Tau & 106         & 19.01   & \nodata        \\
FP Tau & 38           & \nodata & \nodata       \\
GK Tau & 15--35    & 0.14      & 1.244          \\
HL Tau & 43--55     & \nodata & 3.285         \\
RY Tau & 13           & \nodata & 5.242         \\
T Tau  & 41--60      & \nodata & 9.395          \\
UY Aur & \nodata  & \nodata & \nodata         \\

\enddata

\tablenotetext{a}{Compiled from literature by \citet{whit2001}.}
\tablenotetext{b}{Values measured in \cite{calv1998} and compiled by \cite{john2002}.}
\tablenotetext{c}{Compiled from literature by \citet{gude2007}.}

\end{deluxetable}

On VATT, CorMass slit dimensions are 1\ptsec6 in width and 12\ptsec2 in length.  Observations of both the target cTTSs and telluric calibration sources were made using a standard ABBA nod pattern, in which the source was centered in the slit at position A, observed, then `nodded' or moved along the slit to position B, observed twice with exposure times equal to the first exposure, then nodded back to A for the fourth and final exposure.  The nod pattern allows for efficient removal of thermal background, sky emission lines, and dark current.  Data reduction was performed using the IDL program CORMASSTOOL adapted from the Spextool reduction package \citep{cush2004}.  Bad pixel correction, flat fielding, and wavelength calibration are accomplished through the use of the XCORMASSTOOL routine.  The wavelength calibration package utilizes resolved OH telluric emission features to individually calibrate each spectrum.  G2V main sequence stars (HD 26710, HD 283697, HD 56513, and HD 30455) were used as telluric calibration sources.  All H~{\scshape i} Paschen and Brackett series absorption lines not blended with telluric features were removed from the calibrator spectra by linear interpolation prior to performing the telluric correction.  The error introduced from the interpolation procedure is estimated to be less than 1\% of the continuum level.  The corrected target spectra were multiplied by a blackbody function matching the temperature of the calibration sources in order to preserve the shape of the spectral energy distribution.  Figure~1 contains seven multi-epoch observations of DR~Tau, spanning the duration of the survey carried out with CorMass at VATT.

The H~{\scshape i} line fluxes were determined using a standard IRAF Gaussian fits procedure.  The fluxes were then dereddened using A$_{\rm V}$ values from \cite{stro1989}, \cite{keny1995}, and \cite{whit2001} and a standard A$_\lambda$~$\sim$~$\lambda^{-1.84}$ extinction law adopted from \cite{mart1990}.  

\begin{figure}[htbp]
\begin{center}
\includegraphics[angle=90,width=1.0\columnwidth]{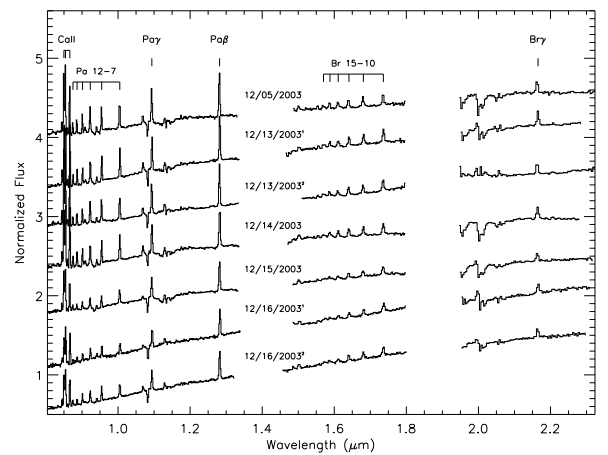}
\caption{As an example of the data collected for the multi-epoch survey, seven of the spectra obtained for the source DR~Tau are presented above.  Each spectrum contains multiple H~{\scshape i} recombination emission features, the Ca {\scshape ii} infrared triplet in emission, and the He~{\scshape i} (1.083 $\mu$m) feature, seen here in absorption.  Individual spectra are labeled with the UT date on which it was observed.  Vertical dotted lines locate the spectral features. }
\label{Fig 1}
\end{center}
\end{figure}

\section{Analysis of H I Line Ratios}
\label{analysis}

The line fluxes for nine Paschen (Pa$\beta$, $\gamma$, $\delta$, 8, 9, 10, 11, 12, and 14) and eight Brackett series (Br$\gamma$, 10, 11, 12, 13, 14, 15, and 16) lines were measured.  These H~{\scshape i} line fluxes\footnote{The large width of the CorMass slit would enable us to very accurately flux calibrate the spectra under photometric observing conditions.  However, some observations were made during inferior conditions introducing error into our flux calibration and line flux measurements.  This discussion of line flux variability refers to the large scale variations attributed to source variability and not the uncertainties in the flux calibration.} appear to differ by greater than an order of magnitude from source to source.  Multi-epoch observations of individual sources reveal significant line variability as can be determined by inspection in Figure~1.

While the magnitudes of the line fluxes and their variability are sensitive to uncertainties in the flux calibration, the line ratios are independent of such errors.  In regimes where the gas is either optically thick or thin in LTE or is well-described by the approximations of hydrogen recombination line theory, these ratios can provide an accurate determination of the physical conditions of the emitting gas.  The spectra in this survey have the advantage over previous studies of H~{\scshape i} line ratios observed for TTS because they allow for a large number of emission lines to be measured simultaneously.  Here we describe the determination and present an analysis of three different series of line ratios and compare these observed values to those predicted by three distinct radiative regimes in an attempt to constrain the physical conditions of the emitting gas.

\subsection{Calculating H I Line Ratios}
\label{HIR}

Figure~2 shows the dereddened line fluxes, including observational uncertainties, for Pa$n_{up}$ to Pa$\beta$ and Br$n_{up}$ to Br$\gamma$.  Only emission lines measured to be 3$\sigma$ detections or better are plotted and only these line are used in this analysis.  Seventy-three of the 104 spectra have at least one useful ratio, and only lines from these spectra are used in this analysis.  None of the observations of RY Tau have 3$\sigma$ detections of two H~{\scshape i} lines, but all other stars in our sample have at least one spectrum in the set of 73.  As shown in Figures 2, a strong linear correlation exists between the H~{\scshape i} line fluxes with surprisingly little scatter, considering that the data represent many different sources and multiple epochs of observations of the same source.  In Figure~3, Pa$\beta$ equivalent widths are plotted versus the values of the Pa$\gamma$/Pa$\beta$ line flux ratios illustrating the intra-source variability and the consistency of the values measured for the line ratios.  Table~2 provides the values for the equivalent widths and flux ratios for the sources designated by colored points in Figure~3 along with the Julian date for each epoch of observation.  For sources such as CY~Tau, BP~Tau, and DR~Tau that possess large epoch-to-epoch variations in the values of their line fluxes and substantially different average flux values, the values of the line ratios remain constant within the uncertainties.  A similarly small scatter in the values of this line ratio is observed for sources such as T~Tau and DG~Tau that possess relatively small variations in the magnitudes of their line fluxes.  
 
The relatively small scatter in the observed values of the line flux ratios for this sample of 15 T~Tauri systems possessing a broad range of intrinsic mass accretion rates (see Table~1) suggests that the ratios are not sensitive to variations in $\dot{M}$.  For this reason, and as a first step toward measuring the average physical conditions of the gas in these systems, we determine a `global' line ratio representative for all of the stars in this survey for each ratio included in the analysis.  In order to determine the representative line ratios, an average value for each ratio weighted by the uncertainty was found.  In addition, the statistical ``estimated variance,'' was calculated to quantify the scatter about the average line ratio values.  In Figure~2, the solid lines have slopes given by the `global' line ratios determined for each pair of line fluxes.  The dashed lines have slopes equal to `global' line ratio plus and minus the estimated variance and thus represent the scatter in the measurements.

Each ratio uses the Paschen or Brackett series transition with the lowest excited state as the reference transition (Pa$\beta$ and Br$\gamma$).  The number of stars with detectable H~{\scshape i} features diminished for the transitions with higher $n$-values.  As a result, the representative line ratios involving the highest transitions in the Paschen and especially the Brackett series are determined by only a few of the TTS in our sample.  However, the small intrinsic source-to-source scatter observed for the other ratios that included larger samples of line fluxes suggest that, even in the cases where only a few sources are used to determine the value of the line ratio, it is reasonable to infer that the ratios are representative of the sample as a whole, within the uncertainties.  We proceed by making a direct comparison between the observed values for these sets of line ratios and the predictions of line formation theory.

 \begin{figure}
 \includegraphics[width=0.9\columnwidth]{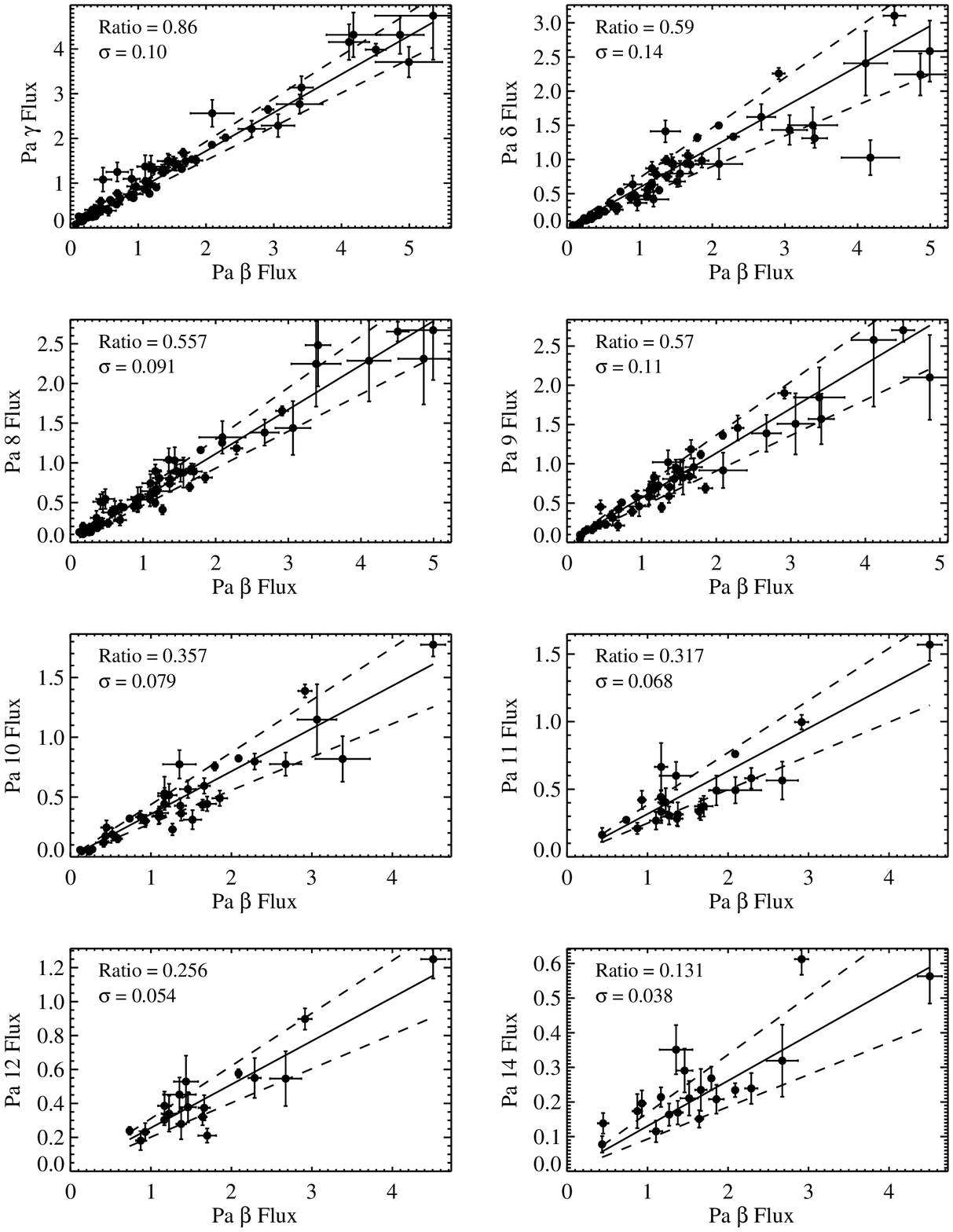}\\
Fig. 2 --- For each spectra in our sample for which we detect these Paschen series lines, we plot the measured Pa$\beta$ line flux versus that measured for eight other Pa$n_{up}$ transitions, with associated measurement uncertainties and find a strong linear correlation for all emission features.  Both axes of all plots are in units of 10$^{-12}$~ergs~cm$^{-2}$~s$^{-1}$.  Each solid line plotted through the data has a slope equal to the weighted average ratio of the data points and represents the `global' line ratio for each plot.  The dashed lines represent the estimated variance and provide a useful measure of the scatter in each of the line ratios shown on each graph.  Also included are similar plots for the Brackett series lines, using Br$\gamma$ and seven other Br$n_{up}$ transitions.
 \end{figure}
 \begin{center}
 \includegraphics[width=0.9\columnwidth]{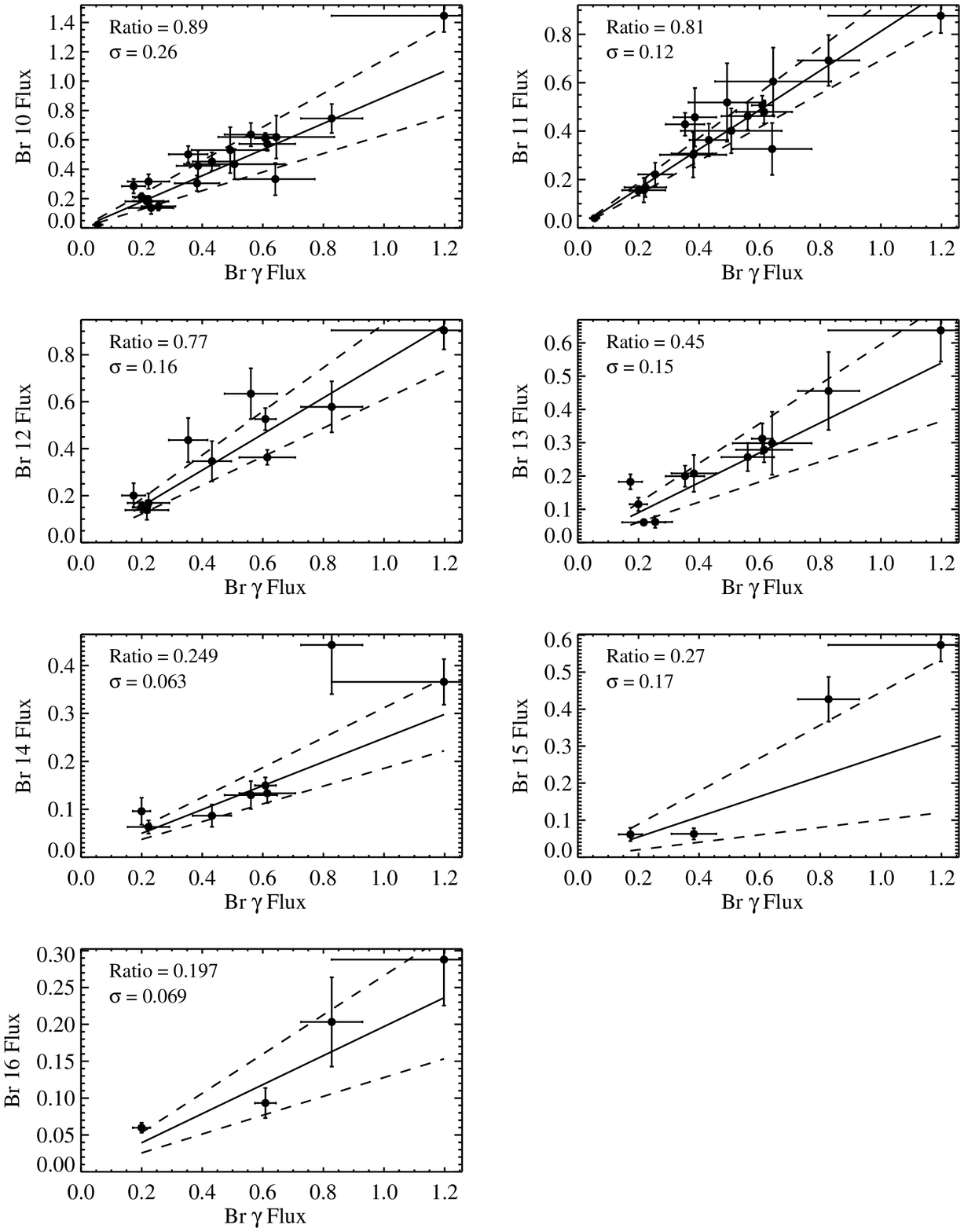}\\
 Fig. 2 --- Continued
 \end{center}



\begin{figure}[htbp]
\begin{center}
\includegraphics[width=1.0\columnwidth]{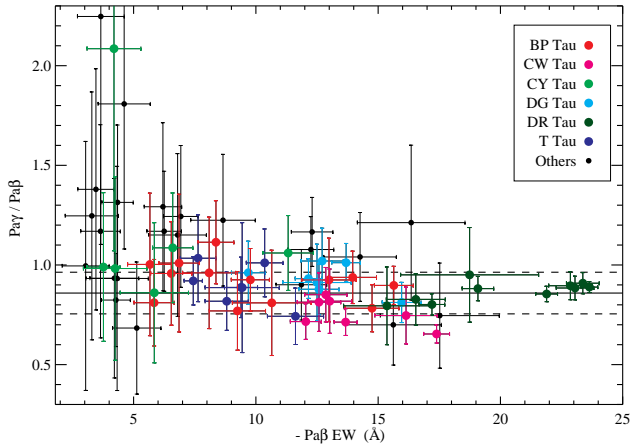}
\caption{For each of the 73 spectra with detectable Pa$\beta$ and Pa$\gamma$ emission features, we plot the values of the Pa$\beta$ equivalent widths versus the values of the Pa$\gamma$/Pa$\beta$ line flux ratios.  Sources for which we have six or more different epochs of observations have been color coded.  The dark horizontal line in the center represents the weighted average of the line ratio for all of the data points.  The dashed horizontal lines represent the ``estimated variance'' of the line ratio value.  For large variations in the magnitude of the Pa$\beta$ EWs, relatively small variations are seen in the values of the line ratio.  The colors highlight this result for the individual sources.}
\label{Fig3}
\end{center}
\end{figure}

\begin{deluxetable}{lcccccc}
\tablewidth{0pt}
\tablecaption{Selected line data.}
\tablehead{
\colhead{Star} &
\colhead{Epoch} & 
\colhead{Pa$\beta$} & 
\colhead{Pa$\gamma$/Pa$\beta$} \\
\colhead{ } &
\colhead{(JD$ - 2450000$)} & 
\colhead{EW(\AA)} & 
\colhead{flux ratio} &
}

\startdata

BPTau
 & 2976.8 & $ -5.67$ $\pm$ 0.91 & 1.00 $\pm$ 0.36 \\
 
 & 2977.8 & $ -6.86$ $\pm$ 0.83 & 1.01 $\pm$ 0.35 \\
 
 & 2986.7 & $ -8.37$ $\pm$ 0.73 & 1.11 $\pm$ 0.21 \\
 
 & 2986.9 & $ -9.2$ $\pm$ 1.1 & 0.77 $\pm$ 0.20 \\
 
 & 2987.7 & $ -6.54$ $\pm$ 0.84 & 0.96 $\pm$ 0.26 \\
 
 & 2988.7 & $ -9.77$ $\pm$ 0.78 & 0.93 $\pm$ 0.16 \\
 
 & 2989.6 & $ -8.1$ $\pm$ 1.1 & 0.96 $\pm$ 0.28 \\
 
 & 2989.9 & $-10.6$ $\pm$ 1.6 & 0.81 $\pm$ 0.26 \\
 
 & 3065.7 & $-13.0$ $\pm$ 1.1 & 0.92 $\pm$ 0.21 \\
 
 & 3074.7 & $ -5.84$ $\pm$ 0.82 & 0.81 $\pm$ 0.22 \\
 
 & 3301.8 & $-14.8$ $\pm$ 1.1 & 0.78 $\pm$ 0.12 \\
 
 & 3399.7 & $-13.96$ $\pm$ 0.97 & 0.94 $\pm$ 0.13 \\
 
 & 3401.8 & $-15.64$ $\pm$ 0.79 & 0.897 $\pm$ 0.096 \\
CWTau
 & 2978.9 & $-13.02$ $\pm$ 0.89 & 0.82 $\pm$ 0.16 \\
 
 & 2986.7 & $-12.04$ $\pm$ 0.63 & 0.716 $\pm$ 0.089 \\
 
 & 2987.6 & $-12.58$ $\pm$ 0.89 & 0.81 $\pm$ 0.15 \\
 
 & 2988.7 & $-13.67$ $\pm$ 0.47 & 0.713 $\pm$ 0.067 \\
 
 & 2989.6 & $-16.1$ $\pm$ 1.3 & 0.75 $\pm$ 0.14 \\
 
 & 3301.8 & $-17.39$ $\pm$ 0.53 & 0.654 $\pm$ 0.045 \\
 
 & 3399.7 & $-12.86$ $\pm$ 0.86 & 0.85 $\pm$ 0.14 \\
CYTau
 & 2978.8 & $ -4.2$ $\pm$ 1.3 & 0.98 $\pm$ 0.46 \\

 & 2988.7 & $ -3.77$ $\pm$ 0.68 & 0.99 $\pm$ 0.37 \\
 
 & 2989.6 & $-11.3$ $\pm$ 1.0 & 1.06 $\pm$ 0.19 \\
 
 & 3074.7 & $ -5.8$ $\pm$ 1.4 & 0.86 $\pm$ 0.35 \\
 
 & 3301.8 & $ -6.60$ $\pm$ 0.82 & 1.09 $\pm$ 0.28 \\
 
 & 3399.7 & $ -4.2$ $\pm$ 1.1 & 2.1 $\pm$ 1.0 \\
DGTau
 & 2978.9 & $-12.71$ $\pm$ 0.86 & 1.02 $\pm$ 0.16 \\
 
 & 2986.9 & $-13.69$ $\pm$ 0.58 & 1.011 $\pm$ 0.096 \\
 
 & 2987.9 & $-15.96$ $\pm$ 0.79 & 0.81 $\pm$ 0.10 \\
 
 & 2988.7 & $-12.15$ $\pm$ 0.54 & 0.931 $\pm$ 0.099 \\
 
 & 2989.7 & $ -9.69$ $\pm$ 0.69 & 0.96 $\pm$ 0.16 \\
 
 & 3301.9 & $-12.58$ $\pm$ 0.82 & 0.88 $\pm$ 0.13 \\
 
 & 3399.8 & $-12.5$ $\pm$ 1.4 & 0.91 $\pm$ 0.19 \\
DRTau
 & 2978.9 & $-23.37$ $\pm$ 0.67 & 0.909 $\pm$ 0.053 \\
 
 & 2986.8 & $-23.64$ $\pm$ 0.34 & 0.889 $\pm$ 0.027 \\
 
 & 2986.9 & $-23.05$ $\pm$ 0.77 & 0.884 $\pm$ 0.059 \\
 
 & 2987.7 & $-22.86$ $\pm$ 0.89 & 0.896 $\pm$ 0.070 \\
 
 & 2988.7 & $-17.20$ $\pm$ 0.52 & 0.802 $\pm$ 0.054 \\
 
 & 2989.7 & $-15.4$ $\pm$ 1.8 & 0.79 $\pm$ 0.19 \\
 
 & 2989.9 & $-16.5$ $\pm$ 1.2 & 0.83 $\pm$ 0.13 \\
 
 & 3074.7 & $-21.89$ $\pm$ 0.45 & 0.854 $\pm$ 0.039 \\
 
 & 3399.6 & $-19.08$ $\pm$ 0.64 & 0.881 $\pm$ 0.061 \\
 
 & 3401.8 & $-18.7$ $\pm$ 2.8 & 0.95 $\pm$ 0.24 \\
TTau
 & 2979.9 & $ -7.44$ $\pm$ 0.37 & 0.92 $\pm$ 0.12 \\
 
 & 2986.9 & $ -7.63$ $\pm$ 0.73 & 1.03 $\pm$ 0.22 \\
 
 & 2987.9 & $-11.6$ $\pm$ 1.1 & 0.74 $\pm$ 0.14 \\
 
 & 2988.7 & $ -9.39$ $\pm$ 0.68 & 0.89 $\pm$ 0.15 \\
 
 & 2989.7 & $ -9.4$ $\pm$ 1.5 & 0.89 $\pm$ 0.33 \\
 
 & 3301.9 & $-10.36$ $\pm$ 0.75 & 1.01 $\pm$ 0.17 \\
 
 & 3401.8 & $ -8.81$ $\pm$ 0.88 & 0.82 $\pm$ 0.15 \\

\enddata

\end{deluxetable}

\subsection{Comparison to Case B models}

\subsubsection{Case B}

The determination of the electron cascade and the population densities of the energy levels in recombining hydrogenic atoms is a well-studied problem that, historically, has been divided into two distinct Cases, A and B \citep{baker1938}.  Case~A applies to a gas for which every transition in the hydrogen atom is optically thin such that any photons emitted by atoms within the gas escape from the emitting region with no subsequent scattering events.  The Case~B approximation of \cite{baker1938} assumed the gas to be optically thick to the ultraviolet photons associated with the Lyman series and continuum and optically thin to photons from all other transitions.  These early models included only radiative capture and cascade and, as a result, were only sensitive to the gas temperature.  In the decades since this work, improvements upon their estimates of the $n, l$-level populations came with the inclusion of collisional effects (i.e., collisional ionization, three-body recombination, and all $\Delta$$n$ transitions; \cite{seat1964,broc1970,broc1971}).  The introduction of collisional effects adds an electron density sensitivity to the relative $n, l$-level populations of the hydrogen atoms.  Therefore, the current state-of-the-art Case~B models \citep{humm1987,stor1995} compute the relative intensities for H~{\scshape i} recombination lines for a range of temperatures (500~$\le$~$T$~$\le$~30000~K) and electron densities (100~$\le$~n$_{\rm e}$~$\le$~10$^{14}$~cm$^{-3}$).

Case~A is applicable to only the lowest density nebula and astrophysical environments and, most likely, will not be the most appropriate approximation for gas located in the very inner regions of an accreting T~Tauri system \citep{oster1989}.  Therefore, in the following analysis, we focus on the Case~B approximation.

\subsubsection{{\rm Pa}$n_{up}$/{\rm Pa}$\beta$, {\rm Br}$n_{up}$/{\rm Br}$\gamma$, and {\rm Br}$\gamma$/{\rm Pa}$n_{up}$}

A Fortran program provided by \cite{stor1995,storc1995} and accompanied by the necessary data files (on-line at http://vizier.u-strasbg.fr/viz-bin/VizieR?-source=VI/64) was used to produce models of the Paschen and Brackett decrements as well as the line ratios of Br$\gamma$ to Pa$n_{up}$ for a grid of 130 different temperature and electron density combinations for 500~$\le$~$T$~$\le$~30,000~K and 100~$\le$~n$_{\rm e}$~$\le$~10$^{14}$~cm$^{-3}$.  For a statistical comparison between the Case~B models and each series of observed line ratios, the reduced Chi-square ($\chi^2$) value for each model was computed.

\begin{figure}[htbp]
\begin{center}
\includegraphics[width=1.0\columnwidth]{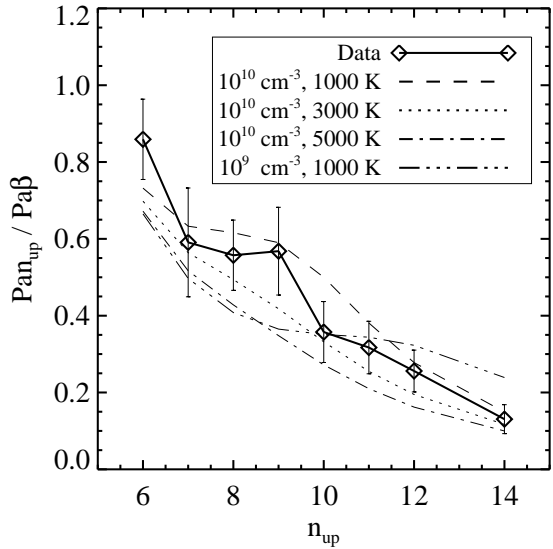}
\caption{Using the weighted mean line ratios for the Paschen series with Pa$\beta$ as the reference transition (see Figure 2a), we plot the observed Paschen decrement curve as a solid line connecting the diamond-shaped data points.  `Scatter bars,' determined from the estimated variance, are plotted atop the data points.  Of the possible 130 computed Case~B models, we overplot the four with the smallest reduced $\chi^2$-values.  The temperatures and electron densities for these models are given in the figure legend.  Reduced $\chi^2$-values can be found in Table~3.}
\label{Fig. 3}
\end{center}
\end{figure}


\begin{deluxetable}{lccc}
\tablewidth{0pt}
\tablecaption{Case B Models with Smallest $\chi^2$}
\tablehead{
     \colhead{Line Ratio}
&   \colhead{n$_{\rm e}$}
&   \colhead{$T$}
&   \colhead{Reduced} \\
     \colhead{}
&   \colhead{(cm$^{-3}$)}
&   \colhead{(K)}
&   \colhead{$\chi^2$} \\
}

\startdata

Pa$n_{\rm up}$/Pa$\beta$  &  $10^{10}$  & $1000$ & 0.93    \\
					  &  $10^{10}$  & $3000$ & 0.99   \\
					  &  $10^{10}$  & $5000$ & 2.35   \\
					  &  $10^{9}$    & $1000$ & 2.81 \\
					  &  $10^{9}$    & $3000$ &  2.83  \\
					  &  $10^{10}$  & $500$   &  2.94  \\
Br$n_{\rm up}$/Br$\gamma$ & $10^{10}$ & ~$500$   &  0.66  \\
					     & $10^{10}$ & $1000$ & 1.48   \\
					     & $10^{9~}$   & $3000$ & 2.19   \\
					     & $10^{9~}$   & $5000$ & 3.09   \\
					     & $10^{10}$ & $3000$ & 3.34   \\
					     & $10^{9~}$   & $1000$ & 3.45   \\
Br$\gamma$/Pa$n_{\rm up}$ & $10^{10}$ & $1000$ &  0.44   \\
					     &  $10^{10}$ & $3000$ &  0.89  \\
					     &  $10^{10}$  & ~$500$ & 1.12  \\
					     &  $10^{9~}$  & $3000$ & 1.68  \\
					     &  $10^{9~}$ & $1000$  & 1.74 \\
					     &  $10^{10}$  & $5000$ & 2.04  \\

\enddata

\end{deluxetable}







In Figure~4, we plot a line ratio curve for the observed Paschen series features using Pa$\beta$ as the reference transition and overplot the four Case~B models with the smallest reduced $\chi^2$-values.  The `scatter bars\footnote{We use the term `scatter bars', instead of error bars, to more accurately reflect the nature of these these vertical bars.  The range of values allowed by each `scatter bar' is not due only to uncertainty in measured values, but also to the actual scatter of the line ratios (i.e., estimated variance) determined for all sources in our sample.}' on each of the data points represent the estimated variance (see $\S\ref{HIR}$) determined from the scatter plots presented in Figure~2a.  The model that best fits the data in Figure~4 (dashed line) has $T$~=~1000~K and n$_{\rm e}$~=~10$^{10}$~cm$^{-3}$.  The parameters and reduced $\chi^2$ values for the best six models are presented in Table~3.  We note that the temperature sensitivity of these models at the best-fit density of 10$^{10}$~cm$^{-3}$ is most apparent at the mid-range $n_{up}$-values where the most separation can be seen between the different models.  In contrast, the temperature is poorly constrained by the Pa$n_{up}$ transitions with the lower and higher $n_{up}$-values.

Figure~5 contains the line ratio curve for the observed Brackett series features using Br$\gamma$ as the reference transition.  Overplotted are four Case~B model fits, all with a density of n$_{\rm e}$~=~10$^{10}$~cm$^{-3}$, which illustrate the temperature sensitivity of the models at the lowest $n_{up}$-values.  The lowest temperature curves for $T$~=~500~K and 1000~K are the two models that most closely fit the data.  The reduced $\chi^2$-values for the six best fit models are presented in Table~3.  Despite the rather large scatter bars associated with these data points, the data significantly rule out the highest temperature Case~B models.  For both the Paschen and Brackett series line ratios, we find that the best Case~B fits to the data have n$_{\rm e}$~=~10$^{10}$~cm$^{-3}$ and $T$~=~1000~K and 500~K for Paschen and Brackett, respectively.  The consistency of the gas temperature and electron density ranges determined by both independent H~{\scshape i} series line ratios and the statistical significance assigned to these fits is remarkable and is strong evidence that the assumptions of the Case~B approximation are valid for the emitting H~{\scshape i} gas.

\begin{figure}[htbp]
\begin{center}
\includegraphics[width=1.0\columnwidth]{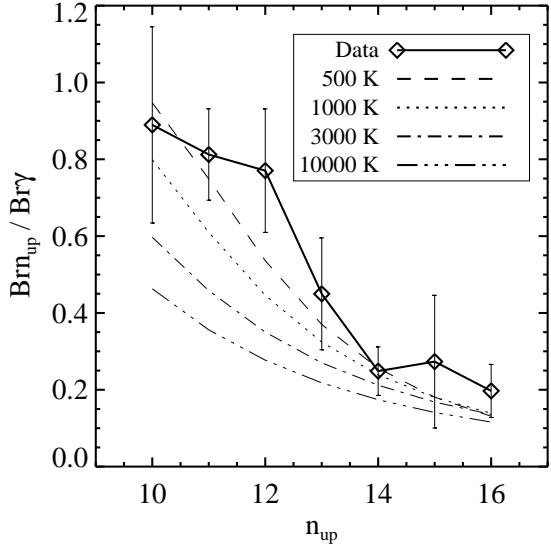}
\caption{Using the weighted mean line ratios for the Brackett series with Br$\gamma$ as the reference transition, we plot the observed Brackett decrement curve as a solid line connecting the diamond-shaped data points (see Figure 2b).  Scatter bars, determined from the estimated variance are plotted atop the data points.  We overplot four Case~B models of constant electron density (n$_{\rm e}$~=10$^{10}$~cm$^{-3}$).  Note that the T~=~500~K (dashed) and 1000~K (dotted) models are the two best-fit models (see reduced $\chi^2$-values in Table~3).}
\label{fig_brntog}
\end{center}
\end{figure}






\begin{figure}[htbp]
\begin{center}
\includegraphics[width=1.0\columnwidth]{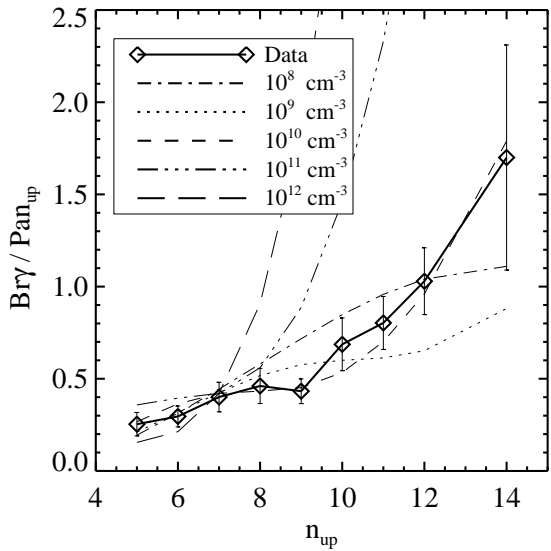}
\caption{The values of the line ratios of Br$\gamma$ to the Pa$n_{\rm up}$ of the Paschen series are plotted.  The broad wavelength coverage between the Brackett and Paschen lines provides a stronger constraint on density as the predicted line ratio curves diverge from the observed curves  significantly at high $n_{up}$.  This sensitivity is demonstrated by plotting four models of constant temperature ($T$~=~1000~K) and separated by one order of magnitude in density along with the data.  The model with the smallest reduced $\chi^2$ has n$_{\rm e}$~=~10$^{10}$~cm$^{-3}$ (see Table~3).}
\label{fig_brgtopan}
\end{center}
\end{figure}






Finally, it is instructive to compare the Case~B model predictions to the observed Br$\gamma$ to Pa$n_{up}$ because of the large wavelength range covered by this ratio.  In Figure~6, we plot the observed line ratio curve for Br$\gamma$ to Pa$n_{up}$ and overplot the best model fit with T~=~10$^3$~K and n$_{\rm e}$~=~10$^{10}$~cm$^{-3}$.  Additionally, four Case~B line ratio curves, all with T~=~10$^3$~K (the temperature of the best model fit) and electron densities that bracket the best-fit density, n$_{\rm e}$~=~10$^{10}$~cm$^{-3}$, are also overplotted.  At low $n_{up}$-values, the differences between the measured and the predicted line ratios are quite small for this temperature.  However, for the line ratios with the largest Pa$n_{\rm up}$ states, the models with the same temperature and different densities are quite distinct from one another, clearly distinguishing the model that most closely matches the data.  As a result, the Br$\gamma$ to Pa$n_{up}$ line ratios place a strong constraint on the n$_{\rm e}$ of the emitting gas.  Once again, the temperature and density of the best-fit Case~B model for Br$\gamma$/Pa$n$ agree well with those determined for both the Paschen and Brackett series ratios.  The large separation in wavelength between the Paschen features and Br$\gamma$ permits this Case~B analysis to be searched for any dependence on the reddening correction.  The agreement between the model temperatures and electron densities determined from all three sets of line ratios suggests that the spectra have been properly corrected for reddening effects.

\subsubsection{$\chi^2$ Contours} \label{sec_contours}

In this section, the reduced $\chi^2$-values for each of the 130
temperature and electron density dependent Case~B models calculated
for the Brackett series and the Br$\gamma$/Pa$n$ line ratios are
combined to determine which models are the best fits to the combined
data\footnote{Note that the $\chi^2$-values for the Paschen series
model comparisons were not included in this calculation to avoid
introducing redundant data and artificially lowering the resulting
$\chi^2$-values.}.  In Figures~7a \& 7b, the reduced $\chi^2$ surface
and contours are plotted, respectively, to illustrate the location of
the models that are the best statistical fits to the data, amongst
the grid of Case~B models.  In Figure~7a, the $\chi^2$-values for
every grid point are presented.  The minimum and surrounding surface
locates the models that best fit the data.  The steep rise of the
surface indicates a sharp increase in the $\chi^2$-values,
highlighting the narrow range of temperatures and electron densities
that fit the data well.  Figure~7b is a contour plot of the confidence
intervals associated with the models with the innermost contour
representing the 60\% confidence interval and the subsequent contours
constrain the models out to the 99.9\% confidence level.  These
confidence intervals also illustrate the tight constraints placed on
both $T$ and n$_{\rm e}$.

Figure~7 illustrates that the Case~B analysis rules out an electron density much larger than 10$^{10}$~cm$^{-3}$, with $10^{11}$~cm$^{-3}$ being completely ruled out.  A density of $\le10^{9}$~cm$^{-3}$ is also ruled out at the 99\% confidence level, for a temperature of 3000~K, and at an even higher confidence level for the rest of the temperatures considered.  The models with densities of $10^{10}$~cm$^{-3}$ have small reduced $\chi^2$ values, making this the most probable density for the emitting gas.  For the gas temperature, a value greater than 5000~K is ruled out at the 99.9\% level.  It is clear that temperatures of $\la 2000$~K are statistically favored, but temperatures lower than 500~K are also allowed and cannot be constrained without models for $T~<~500$~K.  A finer grid of models would be useful for better constraining the range of temperatures and electron densities predicted by the Case~B comparison.  Nonetheless, the analysis as presented here constrains both temperature and density to within a factor of a few.

\begin{figure}
\epsscale{1.0}
\plottwo{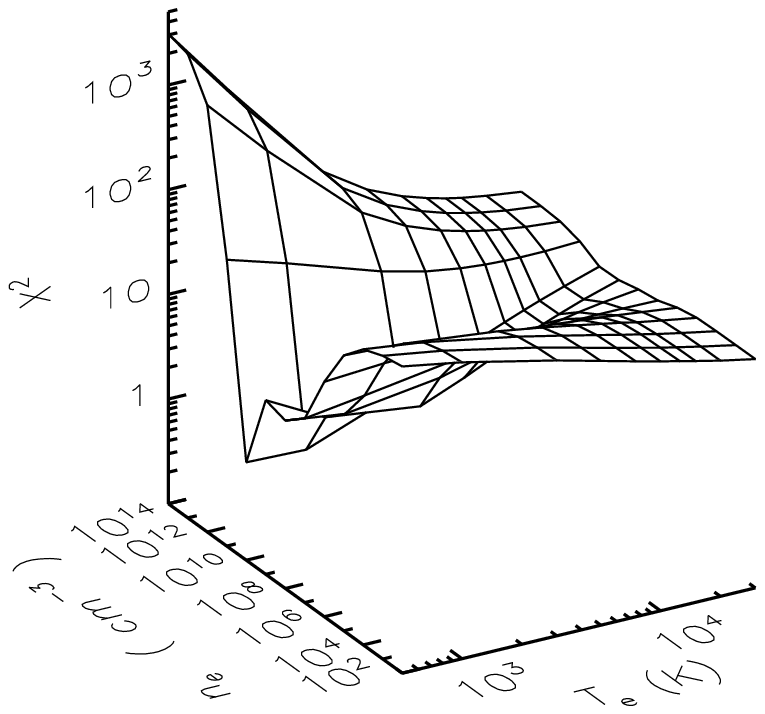}{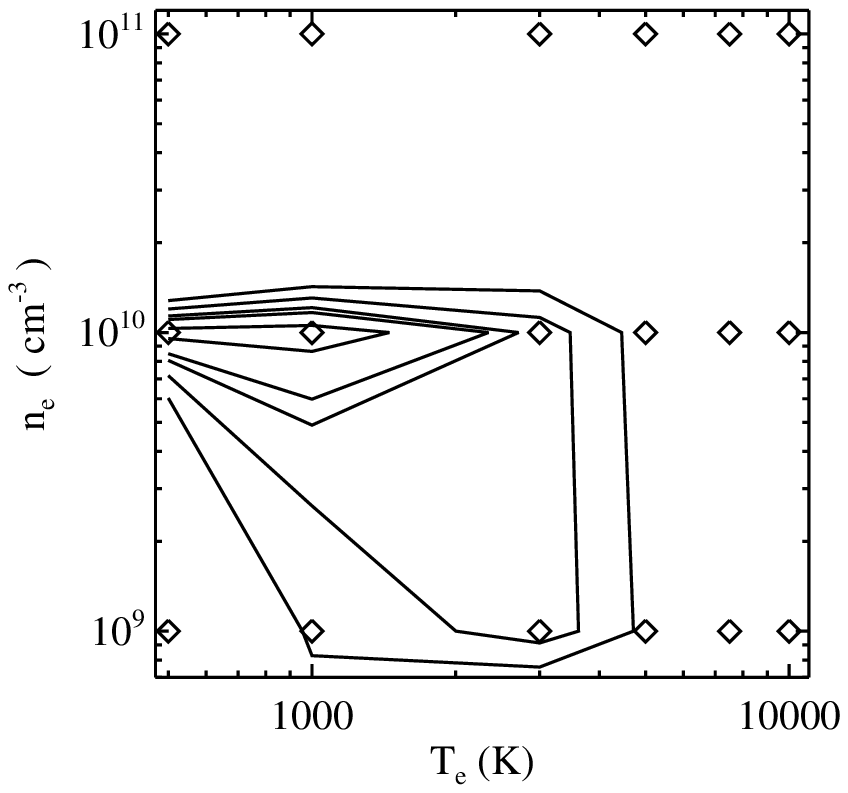}
\caption{a) The reduced $\chi^2$ surface of the 130 density and temperature dependent Case~B models provided by \cite{stor1995} are plotted.  The valley of the surface locates the models with the smallest reduced $\chi^2$ and with its minimum approaching $T~=$~1000~K and n$_{\rm e}$~=~10$^{10}$~cm$^{-3}$.  b) The reduced $\chi^2$ contours are plotted for the values of 1.049, 1.487, 1.666, 2.039, \& 2.513 corresponding to confidence intervals of 60, 90, 95, 99, \& 99.9\% on a $T$ vs. n$_{\rm e}$ plot.  The empty diamonds designate $T$, n$_{\rm e}$ grid points from the 18 Case~B models which fall in the range of the plot.  The contours illustrate the tight constraints placed on densities higher or much lower than 10$^{10}$~cm$^{-3}$ and temperatures greater than 5000~K.}
\label{fig_chisq}
\end{figure}
\subsection{Comparison to Optically Thick and Thin LTE Cases}
\label{lte}

In this section, we compare the observed line ratios to those predicted for gas at a variety of temperatures in both the optically thick ($\tau$~$\gg$~1) or thin ($\tau \ll$~1) LTE regimes.

\subsubsection{Optically Thick Case}

In general, for optically thick gas in LTE, the intensities of all lines approach the value for a blackbody of a single temperature.  Assuming that the line widths are proportional to the wavelength of the line (as expected for Doppler broadening), the line flux ratios are determined by

\begin{eqnarray}
\label{eqn_bbratio}
{F_1 \over F_2} = {{\lambda_1 B_\lambda(\lambda_1,T)} \over 
                   {\lambda_2 B_\lambda(\lambda_2,T)}},
\end{eqnarray}
where $F$ is the line flux, $B_\lambda(\lambda, T)$ is the Planck function, $\lambda$ is the wavelength of the transition, and the subscripts 1 and 2 distinguish between transitions in any ratio.

\begin{figure}
\begin{center}
\includegraphics[width=1.0\columnwidth]{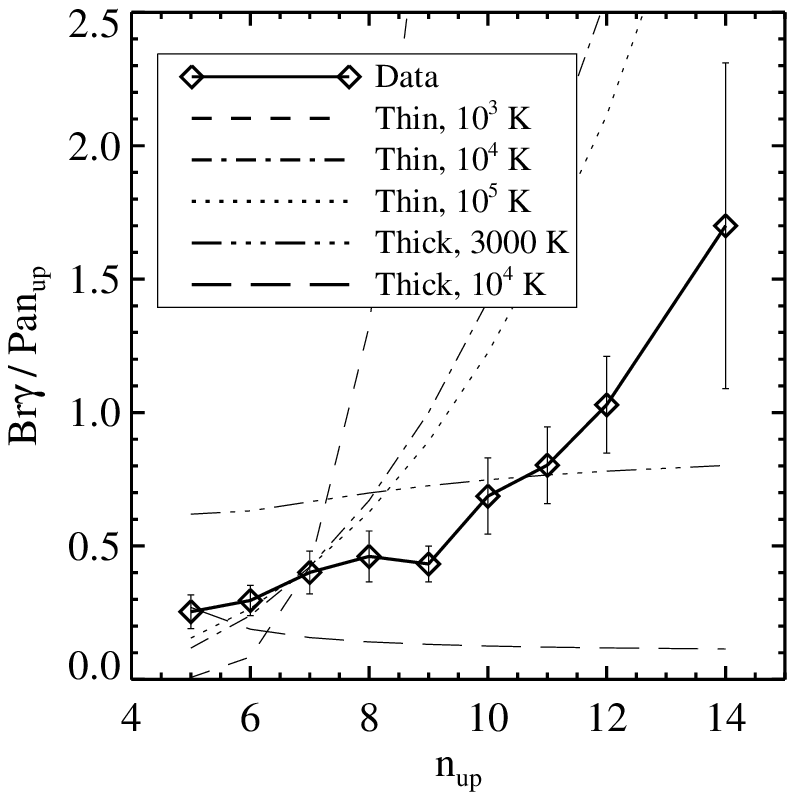}
\caption{The Br$\gamma$/Pa$n$ line ratio curves for gas in both the optically thick and thin LTE regimes are plotted.   A comparison between the observed ratios and those expected for a gas in LTE demonstrates that the emitting gas is not in LTE.  Shown are LTE cases for a variety of temperatures (as indicated in the inset), and there is no temperature that provides a good fit for either LTE case.  The fact that the observed ratio at $n_{up}$~=~7 well-matches the LTE, optically thin value is direct evidence that the two observed lines in this ratio are optically thin.}
\label{fig_lte}
\end{center}
\end{figure}

Figure \ref{fig_lte} compares the prediction of the optically thick LTE case to the observed Br$\gamma$ to Pa$n_{up}$ line ratios.  For the optically thick cases, a blackbody temperature range 3000~$\le$~$T$~$\le$~10$^{4}$~K represents the range of expected temperatures in the accretion flow \citep[e.g.,][]{mart1996, muze2001}.  Figure~\ref{fig_lte} shows that the observed line ratios are not well-matched by the optically thick line ratio curves.  In Table~4, the reduced $\chi^2$-values calculated for several temperatures in the optically thick case, including those plotted in Figure~\ref{fig_lte}, are presented.  A temperature of $\approx$~3800~K is the best statistical fit to the data and has a reduced $\chi^2$ of 5.4, placing it far outside the 99.9\% confidence level.  A similar result is found for the Paschen and Brackett series line ratios (not shown) where the minimum reduced $\chi^2$ values are similarly ruled out.  Therefore, this comparison demonstrates that the optically thick LTE case is not a good approximation of the H~{\scshape i} emitting gas.

\begin{deluxetable}{crc}
\tablewidth{0pt}
\tablecaption{Optically Thick and Thin Reduced $\chi^2$-values }
\tablehead{
\colhead{Case} &
\colhead{$T$} &
\colhead{Reduced}\\
\colhead{} &
\colhead{(K)} &
\colhead{$\chi^2$}
}

\startdata

Thick       & 1000               & 5e6 \\
Thick       & 3000               & 14   \\
Thick       & 3800               &  5.4  \\
Thick       & 10000             & 14   \\
Thin        & 1000               &  2400  \\
Thin        & 10000             & 32     \\
Thin        & 100000           & 18 \\
Thin        & $\infty$~~       & 17 \\

\enddata

\end{deluxetable}

\subsubsection{Optically Thin Case}

We next consider the optically thin LTE case, where the line flux ratios only depend on the relative population densities of the atoms in the upper $n$-levels, the associated Einstein $A$-values, and the wavelengths of the transitions.  For a gas in LTE, the level populations follow a Boltzmann distribution, which was used to calculate the optically thin LTE ratios for this part of the analysis.  In Figure \ref{fig_lte}, the optically thin LTE curves for the Br$\gamma$/Pa$n_{up}$ line ratios with temperatures 1000~$\le$~$T$~$\le$~10$^5$~K are plotted for comparison to the curve representing the observed line ratios. 

As in the optically thick case, the observed data are not fit well by the line ratio values in the optically thin case.  In order to quantify the quality of the fits, the reduced $\chi^2$-values for the optically thin regime were calculated and are presented in Table~4.  For comparison, the models with temperatures of 1000~K, $10^4$~K, and $T$$\rightarrow \infty$ have reduced $\chi^2$-values of $\sim 10^3$, 32, and 17, respectively.  Once again, the LTE case falls outside of the 99.9\% confidence level for any temperature.  Similarly poor correlations for the independent Paschen and Brackett series line ratios also are found for the optically thin case.  Therefore, this comparison demonstrates that the emitting gas is not in the optically thin LTE regime.

By comparing the line ratio curves for the two highest electron density Case~B models (n$_{\rm e}$~=~10$^{11}$ and 10$^{12}$~cm$^{-3}$) in Figure~6 to the optically thin LTE line ratio curves in Figure~8, it is clear that the Case~B values approach those for an optically thin, high density gas in LTE.  The convergence of these two regimes demonstrates how particle collisions dominate the statistics of $n, l$-level populations in the Case~B approximation at high density.  Thus, it is apparent that the emitting H~{\scshape i} gas in the T~Tauri systems in this sample is in a regime where collisions are important for the $n_{up}$-level populations, but not sufficient to give rise to a Boltzmann distribution.  Only a sophisticated treatment of the level populations, such as that found in the current Case~B theory, will properly predict the line ratios.

\subsubsection{Br$\gamma$/Pa$\delta$ Ratio} \label{sec_ratio}

Upon inspection of Figures~6 and~8, we note that the Case~B models and the optically thin LTE case at different temperatures, predict nearly the same value for the Br$\gamma$/Pa$\delta$ line ratio.  This ratio is unique amongst the others in this series because the electrons share the same $n_{up}$-level ($n_{up}$~=~7).  Thus the population density of the upper level of both Br$\gamma$ and Pa$\delta$ is identical, and so this line ratio is independent of the manner in which the upper state is (de)populated and is entirely independent of the temperature and density of the gas\footnote{Note that the individual angular momentum states, $l$, associated with $n_{up}$-state are statistically populated for the densities considered here \citep{broc1971}}.  In the optically thin limit, this line ratio is well-described by the following

\begin{eqnarray}
{F_1 \over F_2} = {{A_1 \lambda_2} \over 
                   {A_2 \lambda_1}},
\label{thin}
\end{eqnarray}

\noindent
where F represents the line flux, $A$ is the Einstein $A$-value, and
$\lambda$ is the wavelength of the transition.  Only optical depth effects can cause the observed ratio to deviate from that predicted by Equation~\ref{thin}.  Therefore, the agreement between the observed value of the Br$\gamma$/Pa$\delta$ line ratio and those predicted by Case~B and the optically thin regime is further evidence that the H~{\scshape i} emission originates in an optically thin gas.

At the same time, the value of the Br$\gamma$/Pa$\delta$ line ratio is accurately
determined from a couple of well-known constants and, therefore, serves as a definitive test of our reddening correction and Case~B analysis.  An error in either the measurement of A$_{\rm v}$, the reddening law, or a systematic error in the determination of the line fluxes would lead to significantly different values measured for any of the line ratios.  Therefore, the agreement between the predicted and measured values of the Br$\gamma$/Pa$\delta$ is a verification of the data analysis techniques, the visual extinction values adopted for these sources, and the reddening law employed.

\section{Discussion}
\label{disc}

The Case~B determination of gas temperature and electron density of the emitting H~{\scshape i} gas is completely independent of any assumptions about where the emission arises (e.g., in a wind, magnetically-channeled accretion flow, or disk).  In this section, we discuss the $T$ and n$_{\rm e}$ ranges constrained by the Case~B analysis in the context of the magnetospheric accretion model, which has had the best success so far in explaining the origin of the H~{\scshape i} emission features.  In this picture, the emitting gas is confined primarily to the accretion funnel and is essentially freely falling onto the star.

\subsection{Accretion Flow Density} 
\label{sec_flowdensity}

For a fiducial accretion rate of $\dot M = 10^{-7} M_\odot$ yr$^{-1}$,
magnetospheric accretion models predict a hydrogen density in the
accretion flow to be $\sim10^{12}-10^{13}$ cm$^{-3}$ \citep{mart1996,
muze1998a, muze2001}.  For the ionization fractions calculated by
\cite{mart1996}, $\sim10^{-2}-10^{-3}$, the range of electron densities for the
accreting gas is expected to be 10$^{9}$-10$^{11}$~cm$^{-3}$.  As shown in section \ref{sec_contours}, the range of electron densities favored by the Case~B analysis falls squarely within the range predicted by the magnetospheric accretion models.  Therefore, the electron densities favored by the Case~B analysis appear to be consistent with the expectations of the magnetospheric accretion models.

However, the tightly constrained values of the electron densities returned
from the Case~B analysis is representative of 15 different stars possessing a
range of spectral types and mass accretion rates.  Given the large range in
the mass accretion rates for the stars in this survey (see Table~1),
which span more than two orders of magnitudes, it is surprising that the range of statistically significant electron densities in the Case~B analysis is at most, a factor of a few (\ref{sec_contours}).

The fact that the large range of mass accretion rates is not reflected
in the range of electron densities appears to require systematic
source-to-source differences in the accretion flows, which serve to
regulate n$_{\rm e}$ of the emitting H~{\scshape i} gas.  In
a steady-state accretion flow, the hydrogen density is
proportional to the mass accretion rate divided by the infall velocity
and divided by the cross-sectional area of the flow.  In the
magnetospheric accretion models, the infall velocity approximately
equals the escape speed from the stellar surface, and this is unlikely
to span more than a factor of a few for the sources in this survey.  
Thus, there remain two potential processes for systematically regulating the
electron density: 1) variations in the cross-sectional areas of the accreting
flows and 2) fluctuations in the ionization fraction.

Measurements of the surface area covered by accretion shocks
\citep[][]{calv1998} have the potential to be reliable probes of the
cross-sectional area of the flows.  \citet{calv1998} determined the areas
of the accretion shock regions on the surfaces of the target stars in this survey
to have a range of more than two orders of magnitude.  Thus, the measured
range in the area of the shocked regions, if representative of the variations in the
cross-sectional areas of the flows, is sufficient to keep the electron density
approximately constant for values of the mass accretion rates, which vary by
a few orders of magnitude.  Furthermore, the results of \citet{calv1998} and
\citet[][and C.\ Johns-Krull, private communication]{vale1993} are consistent with the area of the shocked regions being approximately proportional to $\dot{M}$, a requirement for variations in the cross-sectional area of the flow to regulate the density of the accreting gas.  Although the suggestion that variations in accretion rates are accompanied by nearly equal variations in the cross-sectional areas of the accreting flows is consistent with our finding of relatively constant n$_{\rm e}$, this remains speculative.

In addition to varying the cross-sectional area of the flow as a means of
self-regulating the electron density, fluctuations in the ionization rate due to changes in the mass accretion rate and neutral gas density may also be responsible for maintaining a nearly constant electron density.  Differences in $\dot{M}$ and the gas density may impact the recombination timescale and the ionization faction in such a way as to approximately regulate the electron density in the accreting flow.

Of course, it is unclear whether either of the above processes are responsible
for the small range of electron densities favored by the Case~B analysis of the
sample of TTS in this survey.  Additional observational studies and further
refinements of the line emission models will be necessary to address this issue.

     \subsection{Accretion Flow Temperature and Luminosity} 

\label{sec_flowtemperature}

The preferred values of the gas temperatures derived from the Case~B analysis fall
in the range of $T$~$\la$~2000~K and are substantially lower than the
range of temperatures predicted by the magnetospheric accretion models of
6000~$\la$~$T$~$\la$~12000~K \citep[see \S \ref{intro};][]{muze2001,
mart1996}.  In the models of \cite{muze2001}, the gas temperature was a tunable
parameter.  The ``optimal'' temperatures of $\sim 10^{4}$~K were found, such that
the models reproduced the observed magnitudes of the H~{\scshape i} emission line fluxes.  \cite{mart1996} presented a self-consistent calculation of the thermal structure of the accretion flow, showing that adiabatic compression was the strongest heat source considered for the accreting gas.  That work found that the heating and cooling rates reached equilibrium for a gas temperature of $T~\sim6000$~K, and the total luminosity of the accretion flow is on the order of 10$^{29}$-10$^{30}$~ergs~s$^{-1}$.  This value is comparable to the line luminosities of some of the strongest H~{\scshape i} emission features (e.g., Pa$\beta$, Pa$\gamma$, Br$\gamma$; Figure \ref{Fig3}) observed for sources in this survey (assuming D~$\sim$~140~pc) and cannot account for the total luminosity of the accretion flow.  While a higher gas temperature of $\sim$~10$^4$~K provides a better estimate of the total luminosity of the flow, it requires more heating than that provided by the heat sources included in \cite{mart1996}.  Therefore, the primary mechanism responsible for powering the luminosity of the accretion flows remains unclear.  In addition, the low gas temperatures returned by the Case~B analysis raises questions about the cooling rate of the flow.

          \subsubsection{Cooling Time of Gas in the Accretion Flow}

In order to gain more insight into the thermal structure of the accretion flow, in this section, we estimate the radiative cooling time of the gas, $\tau_{\rm cool}$.  \citet{muze2001} found that the total radiated luminosity of the accretion flow, $L_{\rm af}$, is comprised primarily of hydrogen continuum emission ($\ga 75$\%), plus a significant contribution from emission line fluxes.  In a comparison of $L_{\rm af}$ to the accretion luminosity, $L_{\rm acc}$, assuming that

\begin{eqnarray}
\label{eqn_laf}
L_{\rm af} = f L_{\rm acc} = f {G M_* \dot M \over R_*},
\end{eqnarray}
\noindent
where $M_*$ and $R_*$ are the stellar mass and radius, and $f$ is the fraction of the accretion luminosity radiated away by the accreting gas, \cite{muze2001} found $f$ to be $\sim~5$\%--30\%, for the range of accretion rates exhibited by our
sample stars.

The total mass of material in the accretion flow at any instant in time can be approximated by $\dot M \tau_{\rm ff}$, where $\tau_{\rm ff}$ is the time for material to free-fall from the inner disk truncation radius, $R_{\rm t}$, to the stellar surface.  $\tau_{\rm ff}$ is well approximated by $R_{\rm t}^{3/2} (G M_*)^{-1/2}$, which is several hours for a typical cTTS.  The total thermal energy of the gas in the accretion flow is

\begin{eqnarray}
\label{eqn_uth}
U_{\rm th} \sim {3 \over 2} k T 
                {\dot M \tau_{\rm ff} \over \mu m_{\rm H}},
\end{eqnarray}
where $k$ is the Boltzmann constant, $\mu$ is the mean molecular weight, and $m_{\rm H}$ is the proton mass.

The cooling time, $\tau_{\rm cool}$, is given by, $U_{\rm th}/L_{\rm af}$.  It is instructive to compare the duration of the cooling time with that of the free-fall time by using the ratio

\begin{eqnarray}
\label{eqn_tcool}
{\tau_{\rm cool} \over \tau_{\rm ff}} \sim 0.003       
     \left({0.1 \over f}\right)
     \left({T \over 1000 {\rm K}}\right)
     \left({200~{\rm km~s^{-1}} \over \sqrt{G M_* / R_*}}\right)^2.
\end{eqnarray}

where the input parameters assume $\mu$~=~1.  For the fiducial values in this equation, the cooling time is much shorter than the dynamical (free-fall) time.

This appears to be true for most TTS, since Equation (\ref{eqn_tcool}) is nearly independent of $\dot M$ and $R_{\rm t}$ (the only dependence is via the parameter $f$; \citet{muze2001}).  This result is the consequence of the gas having a luminosity equal to a substantial fraction ($f$) of the gravitational potential energy released (i.e., the accretion luminosity), while having a temperature much cooler than the virial temperature.  Therefore, if the accretion flow is indeed radiating with a luminosity such that $f \sim 0.1$, the free-falling gas has time to reach an equilibrium temperature, set by a balance between heating and radiative cooling.  This will be true, even for the higher temperatures predicted by magnetospheric accretion models.

In order to simultaneously explain the large observed luminosities and low temperatures favored by the Case~B analysis, the accretion flow radiation models need to be improved.  In order to power the emission, additional heating is required than was considered by \citet{mart1996}.  Furthermore, the radiative cooling of the gas must be more efficient than was found by any of the previous models, to explain how the observed line fluxes can be produced by a gas with a lower equilibrium temperature.

          \subsubsection{How Important is Ionizing Radiation?} 

\label{sec_ionizing}

The models of \citet{muze2001} and \citet{mart1996} included an
ionizing source of photons with a luminosity of $\sim10^{28}$ erg
s$^{-1}$ (shortward of 912~\AA).  Since this is not capable of
powering the observed line luminosities, it is not surprising that
these works favored temperatures on the order of $\sim 10^4$ K, where thermal
(collisional) excitation/ionization effects become important for
hydrogen \citep{muze1998a}.  However, the temperature favored by the
Case~B analysis is well-below this temperature, suggesting that a
non-thermal ionization mechanism is at work.

High energy photons generated in both the active coronae and the accretion shocks at the surfaces of active TTS are potential sources of incident radiation capable of ionizing the accreting gas.  The observed X-ray luminosities of the stars in this sample are in the range $10^{29}-10^{31}$ erg s$^{-1}$ (see Table \ref{tab_targets}).  The luminosities in the extreme UV (shortward of 912~\AA) may be similar in strength or potentially even higher \citep{herc2002,berg2003}.  Furthermore, depending on the geometry of the accretion columns and the mass accretion rates, the hydrogen column densities of the accretion flows are estimated to range between $\sim10^{20}-10^{22}$~cm$^{-2}$.  At these column densities, the accreting gas will be capable of absorbing both UV and X-ray photons emitted in the nearby coronae or from accretion shocks on the stellar surface.  The absorption of this radiation, which may be a significant fraction of the high energy photons produced \citep{greg2007} has the potential to power much of the H~{\scshape i} emission.  

A higher incident ionizing flux than that included in previous models should predict a higher ionization fraction for a given temperature.  In turn, this will lead to an increase in the cooling rates from e.g., free-free emission and radiative recombination.  Thus, it qualitatively appears that an irradiated accretion flow model including a realistic ionizing flux may favor a temperature well-below $\sim 10^4$~K and be more consistent with the temperature indicated by the Case~B analysis of this work.

\section{Summary and Conclusions}

In this multi-epoch spectroscopic survey, the line ratios from 17 Paschen and Brackett H~{\scshape i} emission features were measured from a total of 73 spectra of 15 actively accreting cTTS.  The values of the line fluxes measured for the strongest features (e.g., Pa$\beta$) over the entire sample span almost two orders of magnitude.  In spite of this large variation in line fluxes, the measured values of {\em all line ratios} were approximately constant, within the observational uncertainties, from source to source and epoch to epoch.  These results are consistent with earlier work that also found little variation in the value of the Pa$\beta$/Br$\gamma$ ratio \citep{muze2001,natt2006}.  Due to the little variation observed in the values of the line ratios, a single value, given by the uncertainty-weighted average of all the data for each ratio, was adopted. 

     \subsection{Level Populations and Optical Depth}

The measured values for the H~{\scshape i} line ratios and those expected for both the optically thick and thin LTE cases for any temperature were not in agreement.  This result is evidence that the $n, l$-level populations of the H~{\scshape i} gas are {\bf not} in LTE.  On the other hand, the observed ratios were statistically well-fit by
the predictions of the recombination line theory Case~B approximation, tightly constraining the values of $T$ and n$_{\rm e}$.  This is evidence that the characteristic assumptions of Case~B theory hold for the emitting gas.  In addition to the success of the Case~B models, the measured value of the temperature- and density-independent ratio of Br$\gamma$/Pa$\delta$ is consistent with the optically thin value (\S \ref{sec_ratio}).

     \subsection{Density}

The observed line ratios are statistically well-fit by Case~B models
with an electron density of n$_{\rm e}$~$=$~$10^{10}$~cm$^{-3}$ (\S
\ref{sec_contours}).  Furthermore, electron densities of $\ga~10^{11}$~cm$^{-3}$
and $\la~10^{9}$~cm$^{-3}$ are statistically ruled out at the
99\% level or higher (depending on the temperature).  Thus it appears
that the range of electron densities allowed for all 15 stars in this
survey and for all epochs is constrained to lie within only a factor of a
few of $10^{10}$~cm$^{-3}$.

The electron density favored by the Case~B analysis is consistent with
that predicted by magnetospheric accretion models \citep{mart1996,
muze1998a}, for typical cTTS parameters and assuming an ionization fraction of
$\sim10^{-2}$$-$10$^{-3}$ (\S \ref{sec_flowdensity}).  However, it is
a surprise that the allowed range in electron densities is so narrow,
given that the range of accretion rates exhibited by the stars in our
sample spans more than a factor of 100.  If the H~{\scshape i}
emission arises primarily from material in a magnetospheric accretion
flow, this seems to require systematic differences from source to
source that regulate the electron density (see \S\ref{disc}).

     \subsection{Temperature}

The Case~B statistical analysis rules out electron temperatures of $T
\ge 5000$~K for the emitting H~{\scshape i} gas, at the 99.9\% level
(\S \ref{sec_contours}).  Temperatures of $\la 2000$~K are favored,
and temperatures as low as a few hundred Kelvin cannot be
ruled out.  These temperatures are well below the predictions of the
magnetospheric accretion models of $6000 \la T \la 12000$~K \citep{muze2001,
mart1996}.  However, these models currently do not include all important
sources of heating, cooling, and ionization (\S \ref{sec_flowtemperature}).  Therefore,
the lower gas temperatures we find may still be consistent with H~{\scshape i} arising from accreting gas.

Independent of the location of the H~{\scshape i} gas, the Case~B
temperature range appears to be too low to produce the observed emission line
luminosities, unless a non-thermal process is ionizing the gas (\S
\ref{sec_ionizing}).  We propose that a significant flux of ionizing
photons, arising from the stellar coronae and/or accretion shocks near the stellar surfaces, may simultaneously explain the high observed line luminosities and low
temperatures of the emitting H~{\scshape i} gas.

\acknowledgments 

The authors would like to thank Phil Arras, Suzan Edwards, Chris Johns-Krull, James Muzerolle, C. R. O'Dell, Craig Sarazin, Didier Saumon, Keivan Stassun, and David Weintraub for useful discussions about this work.  JSB acknowledges full support through a NSF Astronomy and Astrophysics Postdoctoral Fellowship grant
AST-0507310.  SM was supported by the University of Virginia through a
Levinson/VITA Fellowship, partially funded by The Frank Levinson
Family Foundation through the Peninsula Community Foundation.




\end{document}